\documentclass[conference]{IEEEtran}
\IEEEoverridecommandlockouts
\usepackage{cite}
\usepackage{amsmath,amssymb,amsfonts}
\usepackage{algorithm,algorithmic}
\usepackage{graphicx}
\usepackage{textcomp}
\usepackage{xcolor}
\usepackage{cite}
\usepackage{cite,color}
\usepackage{epsfig}
\usepackage{epstopdf}
\usepackage{url}
\usepackage{color}
\usepackage{multirow}
\usepackage{mathtools}
\usepackage{balance}

\newcommand{\bgeqn}{\begin{equation}}
\newcommand{\edeqn}{\end{equation}}

\newcommand{\beqa}{\begin{eqnarray}}
\newcommand{\eeqa}{\end{eqnarray}}
\newcommand{\beqas}{\begin{eqnarray*}}
\newcommand{\eeqas}{\end{eqnarray*}}

\def\BibTeX{{\rm B\kern-.05em{\sc i\kern-.025em b}\kern-.08em
    T\kern-.1667em\lower.7ex\hbox{E}\kern-.125emX}}

\def\BibTeX{{\rm B\kern-.05em{\sc i\kern-.025em b}\kern-.08em
    T\kern-.1667em\lower.7ex\hbox{E}\kern-.125emX}}

\begin{document}

\title{Two-Stage Optimization for Efficient V2G Coordination in Distribution Power System
}

\author{Pengchao Tian\IEEEauthorrefmark{1}, Siqi Yan\IEEEauthorrefmark{1}, Bikang Pan, Ye Shi\IEEEauthorrefmark{2}\\
% \IEEEauthorblockA{\IEEEauthorrefmark{1}These authors contributed equally to this work}
\thanks{\IEEEauthorrefmark{1}Equal Contribution. \IEEEauthorrefmark{2}Corresponding author.}
\thanks{Pengchao Tian, Siqi Yan, Bikang Pan, Ye Shi are with the School of Information Science and Technology, ShanghaiTech University (email: {\tt tianpch, yansq, panbk2023, shiye@shanghaitech.edu.cn}).}
\thanks{This work was supported by Natural Science Foundation of China  under Grant 62303319, Shanghai Sailing Program under Grant 22YF1428800, and ShanghaiTech AI4S Initiative Project. }
}

% \author{\IEEEauthorblockN{1\textsuperscript{st} Pengchao Tian}
% \IEEEauthorblockA{\textit{School of Information}\\ \textit{and Science Technology}, \\
% \textit{ShanghaiTech University}\\
% Shanghai, China \\
% tianpch@shanghaitech.edu.cn}
% \and
% \IEEEauthorblockN{2\textsuperscript{nd} Ye Shi}
% \IEEEauthorblockA{\textit{School of Information}\\ \textit{and Science Technology}, \\
% \textit{ShanghaiTech University}\\
% Shanghai, China \\
% shiye@shanghaitech.edu.cn}
% \and
% \IEEEauthorblockN{3\textsuperscript{nd} Yuanming Shi}
% \IEEEauthorblockA{\textit{School of Information}\\ \textit{and Science Technology}, \\
% \textit{ShanghaiTech University}\\
% Shanghai, China \\
% shiym@shanghaitech.edu.cn}
% }

\maketitle

\begin{abstract}
With the growing popularity of electric vehicles (EVs), maintaining power grid stability has become a significant challenge. To address this issue, EV scheduling control strategies have been developed to manage vehicle-to-grid (V2G) in coordination with the optimal power flow. In existing studies, such coordination optimization is formulated as a mixed-integer nonlinear programming (MINP), which is computationally challenging due to the binary EV charging and discharging variables. To address this challenge, we develop an efficient two-stage optimization method for this mixed-integer nonlinear coordination problem. This method first employs an efficient technique called the difference of convex (DC) to relax the integrality and reformulate MINP into a series of path-following continuous programming. Although the DC approach shows promising efficiency for solving MINP, it cannot guarantee the feasibility of the solutions. Consequently, we propose a trust region optimization method in stage two that constructs a trust region around DC's solution and then searches for the best feasible solution within this region. Our simulation results demonstrate that, compared to the open-source optimization solver SCIP, our proposed method significantly enhances computational efficiency while achieving near optimality. 
\end{abstract} 


\section{Introduction}

In recent years, the popularity of electric vehicles (EVs) has surged as part of a global call towards clean energy. This trend has posed new challenges for the power grid, including burdening the electricity loads and amplifying the peak electricity demands \cite{khalid2019comprehensive, en13184675}. 
To address this challenge, electric utilities have adopted dynamic pricing and various demand response strategies \cite{muratori2015residential,shi2021distributed}. These initiatives have facilitated the growth of vehicle-to-grid (V2G) technology, which allows for the sale of energy back to the grid during peak times and the charging of EVs during off-peak periods. However, implementing V2G strategies without considering the physical constraints of the power grid can lead to issues such as the clustering effect \cite{shao2012grid,  clement2009impact}. This effect occurs when EVs in a localized area simultaneously charge and discharge, leading to instantaneous power demands that can overload the power grid. To mitigate these risks, it is crucial to coordinate EV scheduling with grid power flow management for the stability and reliability of the power grid \cite{clement2009impact, de2015importance}.

Existing studies have investigated the integration of V2G technology into power grid management. They \cite{kong2022spatial, tsaousoglou2023fair, das2020charging, shi2021model} introduced models that combine the management of EV fleets and the optimal power flow (OPF) schedules, aiming to minimize the cost of grid operations by scheduling generation units, power flow on branches, and the charging/discharging status of EVs. In these existing studies, the coordination of V2G scheduling and OPF can ensure safe operations in the power grid, and they commonly formulate the optimization problem as mixed-integer nonlinear programming (MINP). The integer variables are involved to represent the connection of EVs to the grid or their charging/discharging status. However, the existing MINP solvers like SCIP \cite{achterberg2009scip}, which rely on the branch-and-bound method \cite{lawler1966branch}, will face computational burdens due to their exponential time complexity \cite{basu2021complexity} w.r.t. integer variables. This presents significant challenges in terms of solving speed.

To overcome this challenge, the difference of convex (DC) optimization \cite{tuy1995dc} is a well-known technique when handling nonconvex optimization problems. The principal idea of DC optimization is to reformulate the nonconvex functions into the difference between two convex ones, which is then solved through a path-following optimization method. Applying the DC optimization method to mixed-integer linear programming was first studied in \cite{niu2008dc} and then later extended to MINP in \cite{niu2011efficient}. Combining DC optimization with cutting planes to solve MILP problems was studied in \cite{niu2019parallel}. Recently, Shi et al. \cite{shi2019pmu} developed a path-following computational procedure to optimize the placement of phasor measurement units in smart grids. Although the DC algorithm shows high computational efficiency in solving optimization problems, the integrality relaxation may lead to infeasible solutions to the primal MIP. Consequently, the DC algorithm still cannot ensure safety for hard constraints that are crucial in the power system.

In this paper, we propose an efficient two-stage optimization method for mixed-integer nonlinear V2G scheduling coordinating with OPF. This method aims to enhance computational efficiency while ensuring that the solutions are feasible and near-optimal. In stage one, to address the computational challenges associated with the mixed-integer optimization problem, we employ the DC algorithm to relax the integrality of binary variables and reformulate the primal MINP into a sequence of path-following continuous programs. To ensure the feasibility of the final solutions, we then propose a trust region optimization. The integer variables that satisfy integrality in the DC's solution are treated as a warm-start point, around which we construct a trust region and search for the best feasible solution within the region. We conducted simulation experiments to evaluate the solving speed, feasibility, and optimality gap of our proposed method. The results demonstrate that our method, in comparison to using an open-source solver like SCIP, achieves near-optimal solutions and significantly reduces the solving time. This improvement underscores the effectiveness of our method in handling mixed-integer nonlinear optimization problems in V2G scheduling coordinating with OPF.

\section{Problem Formulation} \label{OPF+EV} 
First, we discuss the OPF in a radial distribution network using the DistFlow model \cite{baran1989network}, and apply the relaxation method from Farivar \cite{farivar2013branch} to address the nonconvex constraints of OPF. Next, we introduce a standard V2G scheduling model that incorporates mixed-integer control variables. As a result, the integration of OPF and V2G scheduling can be modeled as a mixed-integer second-order cone programming problem.

We define ${\cal T} = \{1, \dots, T\}$ as the set of time slots for the optimization period. The set of buses in the power grid is denoted by ${\cal B} = \{1, 2, \dots, N\}$, and ${\cal L}$ represents the set of branches. A branch from bus $i$ to bus $j$ is indicated as $(i,j)\in {\cal L}$. We use $i \rightarrow j$ to signify a directional link from bus $i$ to bus $j$. The subset ${\cal D}\subseteq {\cal B}$ includes buses with connected distributed generators. $\mathcal{V}$ denotes all EVs connected to any charging station, with $\mathcal{V}_i^t \subseteq \mathcal{V}$ specifying the EVs at the charging station on bus $i$ during time slot $t$.

\subsection{Distribution Power Grid Constraints}

The power balance equations ensure that the total power generated and consumed at each bus is balanced, taking into account the power flows to and from connected buses, the power generated at the bus, and the power involved in charging and discharging EVs at the bus. 

We define $R_{i j}$, $X_{i j}$ as the resistance and impedance of the branch between $i$ and $j$. $P_v^{t+}$ and $P_v^{t-}$ denote the charging and discharging power of EV $v$ at time slot $t$; $P_{i,load}^t,  Q_{i,load}^t$ are the active and reactive power load on bus $i$ at time slot $t$.
$P_{i,gen}^t$ and $Q_{i,gen}^t$ represent the active and reactive power generated by the generator; $P_{i j}^t$ and $Q_{i j}^t$ are the active and reactive power flows from bus $i$ to bus $j$. Denote $V_i^t$ as the voltage magnitude at bus $i$ and $I_{ij}^t$ as the current magnitude from bus $i$ to $j$.

The active power balance constraint in the power grid is
\begin{equation}
    \begin{split}
        &P_{j,load}^t+\sum_{k: j \rightarrow k}P_{jk}^t+\sum_{v \in \mathcal{V}_j^t} \left(P_v^{t+}/\eta - P_v^{t-} \eta\right) \\
        &=\sum_{i:i\rightarrow j}\left(P_{ij}^t-R_{ij} (I^t_{ij})^2\right) + P_{j,gen}^t, \forall j \in \mathcal{B}, \label{eq:pbp}
    \end{split}
\end{equation}
where $\eta$ is an efficiency factor during the charging and discharging processes. 
The reactive power balance constraint in the power grid is
\begin{equation}
    \begin{split}
        &Q_{j,load}^t+\sum_{k: j \rightarrow k} Q_{j k}^t\\
        &=\sum_{i: i \rightarrow j}\left(Q_{ij}^t-X_{ij} (I^t_{ij})^2\right)+ Q_{j,gen}^t, \forall j \in \mathcal{B}. \label{eq:pbq}
    \end{split}
\end{equation}
The voltage magnitude in the distribution power grid is
\begin{equation}
\begin{split}
    &{(V_i^t)}^2+\left(R_{i j}^2+X_{i j}^2\right) {(I^t_{i j})}^2
    \\
    &={(V_j^t)}^2+2\left(R_{i j} P_{i j}^t+X_{i j} Q_{i j}^t\right), \forall(i, j) \in \mathcal{L}. \label{eq:branch v}
\end{split}
\end{equation}
The current magnitude in the distribution power grid is
\begin{equation}
    I_{ij}^t = \frac{\sqrt{{(P_{ij}^t)}^2 + {(Q_{ij}^t)}^2}}{V_i^t}, \forall(i, j) \in \mathcal{L}. \label{eq:branch lv}
\end{equation}

Since the constraint (\ref{eq:branch lv}) is nonconvex, which cannot be addressed by existing methods, we relax it into a second-order cone to make the problem solvable \cite{farivar2013branch}. The relaxed inequality constraint is
% \begin{align}
% \left\|\begin{array}{c}
% 2 {P_{ij}^t} \\
% 2 {P_{ij}^t} \\
% {(I_{i j}^t)}^2-{(V_i^t)}^2
% \end{array}\right\|_2 \leq {(I_{i j}^t)}^2+{(V_i^t)}^2. \label{eq:second-order constraint}
% \end{align}
\begin{equation}
    {(P_{ij}^t)}^2 + {(Q_{ij}^t)}^2 \leq (I_{ij}^t)^2(V_i^t)^2, \forall(i, j) \in \mathcal{L}, \label{eq:second-order constraint}
\end{equation}

From the physical properties of the power system, both generated active power and generated reactive power are constrained by the transformer in the charging system:
\begin{align}
    &P_{i,gen}^{min} \leq P_{i,gen}^t \leq P_{i,gen}^{max}, \forall i \in \mathcal{B},  \label{eq:bound Pg}\\
    & Q_{i,gen}^{min} \leq Q_{i,gen}^t \leq Q_{i,gen}^{max}, \forall i \in \mathcal{B}, \label{eq:bound Qg}
\end{align}
where $P_{i,gen}^{min}$ and $P_{i,gen}^{max}$ are the bounds of active generation, $Q_{i,gen}^{min}$ and $Q_{i,gen}^{max}$ are the bounds of reactive generation.
The bounds of bus voltage and branch current are as follows:
\begin{align}
    &\underline{V_i} \leq |V_i^t| \leq \overline{V_i}, \forall i \in \mathcal{B},  \label{eq:bound V}\\
    & -\underline{I_{i j}} \leq I_{i j}^t \leq \overline{I_{i j}}, \forall(i, j) \in \mathcal{L}, \label{eq:bound L}
\end{align}
where $\overline{V_i}, \underline{V_i}$ are the upper and the lower bounds of bus voltage magnitude, $\overline{I_{i j}}, \underline{I_{i j}}$ are the upper and the lower bounds of current magnitude on branches.

\subsection{V2G Constraints}
EVs will arrive at and depart from the charging station with different battery energy. According to existing work \cite{li2022distributionally, li2024distributionally, tang2016model}, the arriving and departing time and battery energy of EVs are assumed to be known. Each EV $v$ should ensure that it reaches $E_v^{dep}$ before departure. 
We define $y_v^{t+},y_v^{t-} \in \{0, 1\}$ as binary decision variables. When $y_v^{t+} = 1$, it represents that EV $v$ is in charging status; and when $y_v^{t-} = 1$, it represents that EV $v$ is in discharging status. If $y_v^{t+} = y_v^{t-} = 0$, then EV $v$ is idle at time slot $t$. To avoid the case that $y_v^{t+} = y_v^{t-} = 1$, we introduce the following constraint:
\begin{equation}
    y_v^{t+} + y_v^{t-} \leq 1, \quad\forall v \in \mathcal{V}.  \label{eq: yvt++yvt-<=1}
\end{equation}
The charging and discharging power are controlled by the binary decision variables within physical bounds:
\begin{align}
    &\underline{P_v} y_v^{t-} \leq P_v^{t-} \leq  \overline{P_v} y_v^{t-}, \quad\forall v \in \mathcal{V}, \label{eq:E-}\\
    &\underline{P_v} y_v^{t+} \leq P_v^{t+} \leq  \overline{P_v} y_v^{t+}, \quad\forall v \in \mathcal{V}, \label{eq:E+}
\end{align}
where $\overline{P_v}, \underline{P_v}$ denote the upper and lower limitation of charging/discharging power. The consistency of battery energy is constrained as
\begin{equation}
    E^t_v = E^{t-1}_v + \left(P_v^{t+} - P_v^{t-}\right) \Delta t, \quad\forall v \in \mathcal{V},  \label{eq:E time}
\end{equation}
where $E^t_v$ represents the battery energy of EV $v$ at time slot $t$, $\Delta t$ is the length of a single time slot. The battery energy should satisfy the physical limitations as
\begin{equation}
    \underline{E_v} \leq E_v^t \leq \overline{E_v}, \quad\forall v \in \mathcal{V},  \label{eq:E bound}
\end{equation}
where $\overline{E_v}, \underline{E_v}$ are the upper and the lower bounds of battery energy.

\subsection{Overall Optimization Problem}
The objective of the optimization problem is to minimize the total system cost, which includes the costs associated with power generation by generators and EVs at charging stations. The cost function is defined as
\begin{equation}
\label{eq: obj}
    C = \sum_{t\in\mathcal{T}} \left(\sum_{j \in \mathcal{D}} c(P_{j,gen}^t) + \beta^t \sum_{v\in \mathcal{V}} \left(P_v^{t+} - P_v^{t-}\right) \right),
\end{equation}
where $\beta^t$ is the electricity price at time slot $t$, the function $c(P_{j,gen}^t)$ represents the cost of power generation, which can be either linear or quadratic in MatPower \cite{zimmerman2010matpower}. 
The overall optimization problem is defined as
\begin{equation}
\begin{gathered}
\label{eq:primal}
\min_{\mathcal{X}, \mathcal{Y}} \sum_{t\in\mathcal{T}} \left(\sum_{j \in \mathcal{D}} c(P_{j,gen}^t) + \beta^t \sum_{v\in \mathcal{V}} \left(P_v^{t+} - P_v^{t-}\right) \right) \\
\text{s.t.} \quad (\ref{eq:pbp})-(\ref{eq:branch v}), (\ref{eq:second-order constraint})-(\ref{eq:E bound}),
\end{gathered}
\end{equation}
where $\mathcal{X}$ is the set of continuous control variables $\{V_j^t, I_{i j}^t, P_{i j}^t, Q_{i j}^t, P_{j,gen}^t, Q_{j,gen}^t, P_v^{t+}, P_v^{t-}\}$ and $\mathcal{Y}$ is the set of the binary control variables $\{y_{v}^{t+}, y_{v}^{t-}\}$. 
Due to the integer variables $\mathcal{Y}$ and the second-order cone constraint (\ref{eq:second-order constraint}), the primal optimization problem (\ref{eq:primal}) is a MISOCP. 

% \red{The computational challenge mainly arises from integer variables, and the existing solvers generally require to build an exponential binary search tree to traverse the values of integer variables, so as to find the optimal solution or feasible solution. Empirically, we can use the DC algorithm to heuristically find a point near the optimal solution, and then search for the best feasible solution in the confidence domain. This limits the number of permutations and combinations of integer values and greatly reduces the worst-case time complexity.}

% The computational challenge primarily stems from integer variables. Current solvers typically necessitate constructing an exponential binary search tree to explore the integer variable values in order to discover the optimal or feasible solution. Empirically, the DC algorithm can be employed to heuristically find a point near the optimum. Subsequently, the best feasible solution is searched within the trust region. This approach restricts the range of permutations and combinations of integer values, leading to a significant reduction in the worst-case time complexity.

\section{Efficient Two-Stage Optimization Method}
In this section, we detail our two-stage method to address the computational challenges in MISOCP for V2G scheduling and OPF control while ensuring safe operations in the power grid. The computational challenge primarily stems from integer variables. Current solvers typically necessitate constructing an exponential binary search tree to explore the integer variable values in order to discover the optimal or feasible solution. We first employ the DC algorithm to relax the nonconvex integrality constraints of the MISOCP, transforming the problem into a series of path-following SOCP problems. This step significantly reduces the computational complexity by converting mixed-integer program into a continuous one, which is easier to solve using conventional convex optimization solver. In stage two, to ensure that the solutions are feasible and of high quality, we implement trust region optimization. This method uses the solution from the DC algorithm as a warm-start point and constructs a trust region around the point, within which searches for the best feasible solution.

% \subsection{Presolve solution via difference of convex}
\subsection{Stage-1: DC Algorithm for Integer Variables}
To tackle binary control variables $\mathcal{Y}$ represented as vector form $\mathbf{y}$, we design a DC algorithm to relax the integrality. Traditionally, MINP is solved by the branch-and-bound algorithm through a binary search tree, which requires exponential complexity. The DC algorithm reformulates such MISOCP into path-following SOCP problems, as a consequence, the solving speed will be improved. One can see that the following equation always holds: 
\begin{equation}
    \begin{split}
        \mathbf{y}\in \{0, 1\}^n \equiv \{\mathbf{y} : \mathbf{y} - \mathbf{y} \circ \mathbf{y} \leq 0, \mathbf{y}\in [0, 1]^n\},
    \end{split} 
    \label{DC reformulation}
\end{equation}
where $\circ$ is the element-wise product between vectors.

Accordingly, we can relax binary constraints into continuous constraints. Define a difference of convex function as $g(\mathbf{y}) = \mathbf{y} - \mathbf{y} \circ \mathbf{y}$ and an equivalent problem is reformulated as 
\begin{equation}
\begin{gathered}
\label{L-relax}
\min_{\mathcal{X}, \mathcal{Y}} \sum_{t\in\mathcal{T}} \left(\sum_{j \in \mathcal{D}} c(P_{j,gen}^t) + \beta^t \sum_{v\in \mathcal{V}} \left(P_v^{t+} - P_v^{t-}\right) \right) \\
\text{s.t.} \quad (\ref{eq:pbp})-(\ref{eq:branch v}), (\ref{eq:second-order constraint})-(\ref{eq:E bound}),\\
g(\mathbf{y}) \leq 0, \quad \mathbf{y} \in [0, 1]^n. \\
\end{gathered}
\end{equation}
Note that the term $-\mathbf{y} \circ \mathbf{y}$ in $g(\mathbf{y})$ is non-convex. A well-designed path-following computational procedure can handle this problem. Specifically, we first develop the lower bounding approximation for the convex term $\mathbf{y} \circ \mathbf{y}$. At the $t$-th iteration, the current point is denoted as $\mathbf{y}^{(t)}$ and the algorithm searches for the next point. According to the first-order Taylor expansion at $\mathbf{y}^{(t)}$, the following inequality holds true:
\begin{equation}
    \begin{split}
        \mathbf{y} \circ \mathbf{y} &\geq \mathbf{y}^{(t)} \circ \mathbf{y}^{(t)} +  \nabla (\mathbf{y}^{(t)} \circ \mathbf{y}^{(t)}) \circ (\mathbf{y}-\mathbf{y}^{(t)})\\
        &= 2\mathbf{y}^{(t)} \circ \mathbf{y} - \mathbf{y}^{(t)} \circ \mathbf{y}^{(t)}.
    \end{split}
\end{equation}
Therefore, an upper bound approximation at $\mathbf{y}^{(t)}$ for $g(\mathbf{y})$ can be easily obtained in the following form:
\begin{equation}
    \begin{split}
        g(\mathbf{y}) &\leq g^{(t)}(\mathbf{y}) \\
        &:= (1 - 2\mathbf{y}^{(t)}) \circ \mathbf{y} + \mathbf{y}^{(t)} \circ \mathbf{y}^{(t)},
    \end{split}
    \label{DCfunc}
\end{equation}
which is a linear function of $\mathbf{y}$.

At the $t$-th iteration, the following SOCP is solved to generate the next iterative point $\mathbf{y}^{(t+1)}$:
\begin{equation}
    \begin{gathered}
    \label{path-follow}
    \min_{\mathcal{X}, \mathcal{Y}} \sum_{t\in\mathcal{T}} \left(\sum_{j \in \mathcal{D}} c(P_{j,gen}^t) + \beta^t \sum_{v\in \mathcal{V}} \left(P_v^{t+} - P_v^{t-}\right) \right) + \lambda g^{(t)}(\mathbf{y}) \\
    \text{s.t.} \quad (\ref{eq:pbp})-(\ref{eq:branch v}), (\ref{eq:second-order constraint})-(\ref{eq:E bound}),\\
    \mathbf{y} \in [0, 1]^n, \\
    \end{gathered}
\end{equation}
where $\lambda \in \mathbb{R}$ is the coefficient of the penalty term. 

Algorithm \ref{algo: DCA} demonstrates the details of the DC algorithm. Due to the relaxation of integer variables, the algorithm can obtain the solution $\hat{\mathbf{x}}, \hat{\mathbf{y}}$ very efficiently and its objective value is close to the optimum. However, only partial elements in $\hat{\mathbf{y}}$ may satisfy the integer constraint and the other elements are still decimal, which means $\hat{\mathbf{y}}$ is infeasible to the primal problem. To address this challenge and ensure feasibility, we then propose the trust region optimization in the next stage.
% we will construct a trust region around the $\hat{\mathbf{y}}$ and search for the best feasible solution.

\begin{algorithm}[!htp]
 \caption{The DC Algorithm for Integer Variables}
    \label{algo: DCA}
    \begin{algorithmic}[1] % 控制是否有序号
        \renewcommand{\algorithmicrequire}{\textbf{Parameter:}}
        \REQUIRE The coefficient of the penalty term $\lambda$;
        \renewcommand{\algorithmicrequire}{\textbf{Input:}}
        \REQUIRE  Primal MISOCP instance $\mathcal{P}$ (\ref{eq:primal}); 
	    \renewcommand{\algorithmicensure}{\textbf{Output:}}   
        \ENSURE The DC outcome $\hat{\mathbf{x}}, \hat{\mathbf{y}}$; 
        \STATE Initialize $\mathbf{y}^{(0)}$, $t=0$;
        \STATE Relax the integrality and formulate SOCP (\ref{path-follow});
        \WHILE{$\max_i|y_i - \operatorname{round}(y_i)| \geq 10^{-5}$ and $t<L$}
        \STATE $\mathbf{x}^{(t+1)}, \mathbf{y}^{(t+1)} \leftarrow$ Solution of problem (\ref{path-follow});
        \STATE $t \leftarrow t + 1$;
        \ENDWHILE
        \STATE $\hat{\mathbf{x}}, \hat{\mathbf{y}} \leftarrow \mathbf{x}^{(t)}, \mathbf{y}^{(t)}$;
        \STATE \textbf{return} $\hat{\mathbf{x}}, \hat{\mathbf{y}}$.
    \end{algorithmic}
\end{algorithm}

\subsection{Stage-2: Trust Region Optimization} 
Since the integrality relaxation is applied in the DC algorithm, there are constraint violations of the DC's outcomes. As reliability and security issues are crucial in power systems, although the DC approach shows promising efficiency for solving MINP, it is not practical in power control. As the observation in numerical experiments that most elements of DC's outcome satisfy integrality, thus a naive approach is to fix all integer values and optimize a new smaller sub-MINP of the remaining variables in $\mathcal{Y}$ to find a solution with no constraint violations in the primal problem. However, directly fixing integer values may sometimes result in no feasible domain remaining in the sub-MINP problem. Consequently, we propose a trust region optimization method, which constructs a trust region around the warm-start point given by DC's outcomes and searches for the best feasible solution. 

We define $\mathcal{S}_0$ as the set of indices for elements in DC's outcome $\hat{\mathbf{y}}$ that are close to 0, and $\mathcal{S}_1$ as the set of indices for those are close to 1. We then create a warm-start point $\bar{\mathbf{y}}_{\mathcal{S}}$ through these elements, where $\bar{y}_{i} = 0$ if $i \in \mathcal{S}_0$ and $\bar{y}_{i} = 1$ if $i \in \mathcal{S}_1$.  Denote the set $\mathcal{S} = \mathcal{S}_0 \cup \mathcal{S}_1$ and $\mathbf{y}_\mathcal{S}$ as the variables in $\mathbf{y}$ with indices in $\mathcal{S}$, the naive fixing approach is to solve the problem with a new constraint as: 
\begin{equation}
    \mathbf{y}_\mathcal{S} = \bar{\mathbf{y}}_{\mathcal{S}}.
\end{equation}
However, the discrepancies between $\bar{\mathbf{y}}_\mathcal{S}$ and the optimal solution $\mathbf{y}^\star_\mathcal{S}$ will lead to suboptimal or even infeasible outcomes.

\begin{algorithm}[!htp]
 \caption{Efficient Two-Stage Optimization Method}
    \label{alg: ols}
    \begin{algorithmic}[1] % 控制是否有序号
        \renewcommand{\algorithmicrequire}{\textbf{Parameter:}}
        \REQUIRE The radius $\Delta$ and a nonnegative value $\lambda$;
        \renewcommand{\algorithmicrequire}{\textbf{Input:}}
        \REQUIRE  Primal MISOCP instance $\mathcal{P}$ (\ref{eq:primal}); 
	    \renewcommand{\algorithmicensure}{\textbf{Output:}}   
        \ENSURE The solution $\mathcal{X, Y}$; 
        \STATE Compute $\hat{\mathbf{y}}$ through DC Algorithm \ref{algo: DCA};
        % \STATE Initialize an empty expression list $E$
        \STATE Check integrality to obtain $\mathcal{S}$;
        % and $\bar{\mathbf{y}}_{\mathcal{S}}$;
        \FOR{$i \in \mathcal{S}$}
        \STATE Create auxiliary binary variable $\delta_i$;
        \IF{$i \in \mathcal{S}_0$}
        \STATE Create constraint $y_i \leq \delta_i$;
        \ELSIF{$i \in \mathcal{S}_1$}
        \STATE Create constraint $1 - y_i \leq \delta_i$;
        \ENDIF
        \ENDFOR
        \STATE Create new constraint $\sum_{i\in\mathcal{S}} \delta_i \leq \Delta$;
        \STATE Let $\mathcal{P}^\prime$ denote the sub-problem (\ref{eq: search}) with new constraints and variables;
        \STATE $\mathcal{X, Y} \leftarrow$ Call MINP solver for sub-problem $\mathcal{P}^\prime$;
        \STATE \textbf{return} $\mathcal{X, Y}$.
    \end{algorithmic}
\end{algorithm}

In contrast, our method treats $\bar{\mathbf{y}}_\mathcal{S}$ as a warm-start point and constructs a trust region around it. The new domain is the intersection of the primal feasible domain and a sphere centered at the warm-start point. A properly defined trust region can ensure a feasible domain exists in the new sub-MINP. Optimizing within this smaller domain is generally faster than solving the primal problem. Moreover, given that $\bar{\mathbf{y}}_{\mathcal{S}}$ closely approximates $\mathbf{y}^\star_{\mathcal{S}}$, we can effectively search for a near-optimal solution with the following  trust region constraint:
\begin{equation}
\| \mathbf{y}_{\mathcal{S}} - \bar{\mathbf{y}}_{\mathcal{S}} \|_1 \leq \Delta, 
\end{equation}
where $\Delta$ is a hyperparameter to represent the size of the trust region. Our method leverages the DC algorithm to efficiently reduce the scale of the sub-MINP problem. Consequently, we improve the solving speed by optimizing the sub-problem with respect to $\bar{\mathbf{y}}_\mathcal{S}$ as below:
\begin{equation}
\label{eq: search}
\begin{gathered}
\min_{\mathcal{X}, \mathcal{Y}} \sum_{t\in \mathcal{T}}\left(\sum_{j \in \mathcal{D}} c\left(P_{j, gen}^t\right)+\beta^t \sum_{v \in \mathcal{V}} \left(P_v^{t+} - P_v^{t-} \right)\right) \\
% \text { s.t. (2)-(4),(6)-(13),(15) }
 \begin{aligned}
    \text{s.t.} \quad &(\ref{eq:pbp})-(\ref{eq:branch v}), (\ref{eq:second-order constraint})-(\ref{eq:E bound}),\\
    & \| \mathbf{y}_{\mathcal{S}} - \bar{\mathbf{y}}_{\mathcal{S}} \|_1 \leq \Delta.
\end{aligned}
\end{gathered}
\end{equation}

Note that, if $\Delta = 0$, the strategy is the same as rounding and fixing variables $\bar{\mathbf{y}}$ into $\{0, 1\}$. The trust region method consistently achieves better objective values by avoiding the incorrect assignment of the fixing strategy, which can lead to suboptimal or infeasible solutions. Section \ref{sec: simulation} will present computational experiments that further validate this advantage. The execution details of our two-stage method are outlined in Algorithm \ref{alg: ols}. After computing the DC's outcome, we select certain elements from $\hat{\mathbf{y}}$ to serve as the center point $\bar{\mathbf{y}}_{\mathcal{S}}$ of the trust region. We then optimize the best feasible solution of sub-MINP that ensures safe operation in the power grid.

% \red{In The coordination of V2G control with the power grid, a large number of integer variables will cause the calculation amount to be unable to meet the needs of dynamic changes of the power grid. In the experimental results, we will see that the two-stage method greatly reduces the computation time and achieves a near-optimal solution. This method has the potential to meet the system requirements in real time in large-scale EV scenarios.}

In the integration of V2G control with the power grid, a significant quantity of integer variables can lead to computational burdens that may not adequately adapt to the dynamic requirements within the power grid. Through the implementation of our two-stage approach, grid systems can enhance their ability to provide solutions more effectively. Experimental results will demonstrate that the two-stage approach substantially reduces the solving time and achieves a near-optimal solution. These solutions hold promise for addressing real-time demands in large-scale control scenarios.

% \red{In practical applications, the coordination of V2G control with the power system must address the need for reduced response times. Moreover, the increasing integration of large numbers of EVs into the grid poses a significant surge in computational complexities. Through the implementation of our two-stage approach, grid systems can enhance their ability to provide solutions more effectively. These solutions hold promise for addressing real-time demands in large-scale control scenarios.}

\section{Simulation} \label{sec: simulation}

\subsection{Simulation Setup}
We conduct experiments on the radial distribution power networks Case 18 and Case 69 with structures, physical limits, and the cost function $c\left(P_{j, gen}^t\right)$ provided by Matpower \cite{zimmerman2010matpower}. The time horizon $T$ is set to 24, with each time slot representing an hour. The solver used for the primal MISOCP problem and the trust region optimization is SCIP \cite{achterberg2009scip}, integrated into the Python environment through CVXPY \cite{diamond2016cvxpy, agrawal2018rewriting}. All experiments are conducted on a processor with 2 Intel Xeon 5218R CPUs (2.1GHz, 20 cores).

\begin{figure}[!hb]
    \centering
    \includegraphics[scale = 0.37]{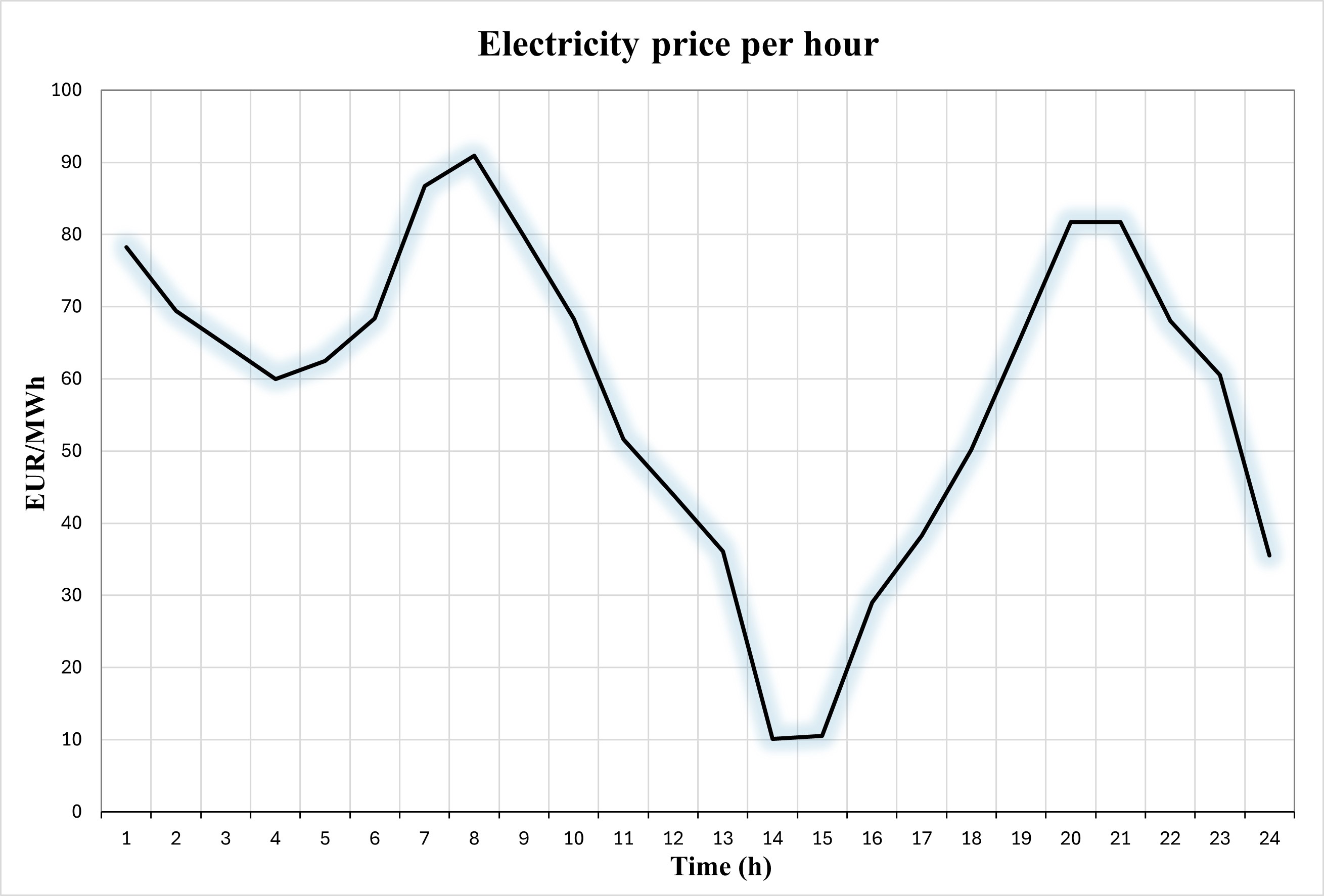}
    \caption{Electricity price in a day.}
    \label{fig: price}
\end{figure}

\begin{table*}[h!]
\begin{center}
\caption{Performance of Our Two-Stage Method}
\label{tab: optimality and efficiency}
\begin{tabular}{cccc cccc}
\hline
 & \multicolumn{3}{c}{Case 18} & & \multicolumn{3}{c}{Case 69}\\
\cline{2-4} \cline{6-8} & Optimum & Ours & Gap (\%) & & Optimum & Ours & Gap (\%)\\
\hline
% \hline
Obj.        & 3070.656   &   3070.734 & 0.019 & & 2287.470 & 2287.525 & 0.007\\
$C_{ev}$    & 423.640    &   423.604  & 0.040 & & 825.964 & 826.177 & 0.026\\
$C_g$       & 2647.017   &   2647.130 & 0.008 & & 1461.506 & 1461.349 & 0.011\\
Time (s)    & 1141.378   &   17.578   & 98.460 & & 8979.523 & 89.939 & 98.998\\
\hline 
\end{tabular}
\end{center}
\end{table*}

In simulations, the charging station is positioned at bus 6 and the number of EVs connecting to the grid varies randomly between 1 and 5 per time slot. The state of charge for EVs arriving ranges from 20\% to 40\%, and for those departing, it should be fully charged. Each EV has a battery capacity of 100 kWh. Charging/discharging power ranges between 10 kW and 20 kW. The efficiency factor $\eta$ of charging and discharging is 0.8. The grid's structure and limits are defined according to Matpower standards \cite{zimmerman2010matpower}. To increase the variability of the experimental data, Gaussian noises with a mean of 0 and variances of 1 and 0.1 squared are added to the load on each bus and to the impedance and reactance on each branch, respectively. The price of electricity at the charging stations, which includes Gaussian noise with a variance of 0.5 squared, is based on the real hourly electricity prices from April 9, 2024, in Germany \cite{online}. The electricity price in a day is shown in Figure \ref{fig: price}. There are about 100 data samples generated for each case in our experiments.

We conduct comprehensive experiments to evaluate the feasibility, optimality gap and solving speed of our proposed two-stage method. To compare the effectiveness of our proposed method with the fixing strategy, we compute the feasibility ratio across our data samples as $N_{f}/N$,
where $N$ is the number of data samples in our study and $N_{f}$ is the number of instances that obtain feasible solutions. Note that we can adjust the value of $\Delta$ to regulate the size of the trust region, increasing $\Delta$ can improve the Feasibility ratio and reduce the gap with the optimal solution.

To assess the quality of our two-stage method in computing solutions, we employ the optimality gap metric. This metric quantifies the difference between the solution $\tilde{x}$ obtained by our method and the global optimum $x^\star$. It effectively measures how close our solution is to the best possible solution:
\begin{equation}
    \operatorname{Gap}(\tilde{x}) = \frac{|\tilde{x} - x^\star|}{|x^\star| + \epsilon} \times 100\%,
    \label{gap}
\end{equation}
where $\epsilon = 10^{-8}$ is introduced to prevent division by zero. The objective function $C$ consists of the cost of generators
\begin{equation}
    C_g = \sum_{t\in\mathcal{T}} \sum_{j \in \mathcal{D}} c(P_{j,gen}^t),
\end{equation}
and the electricity cost for EVs
\begin{equation}
    C_{ev} = \sum_{t\in\mathcal{T}} \beta^t \sum_{v\in \mathcal{V}} (P_v^{t+} - P_v^{t-}).
\end{equation}
We calculated the value and optimality gap of each component. Moreover, the processing time is reported to highlight our method's efficiency in solving these problems.

\subsection{Simulation Results}
\subsubsection{Optimality Gap and Efficiency} 
\begin{figure}[!hb]
    \centering
    \includegraphics[scale = 0.4]{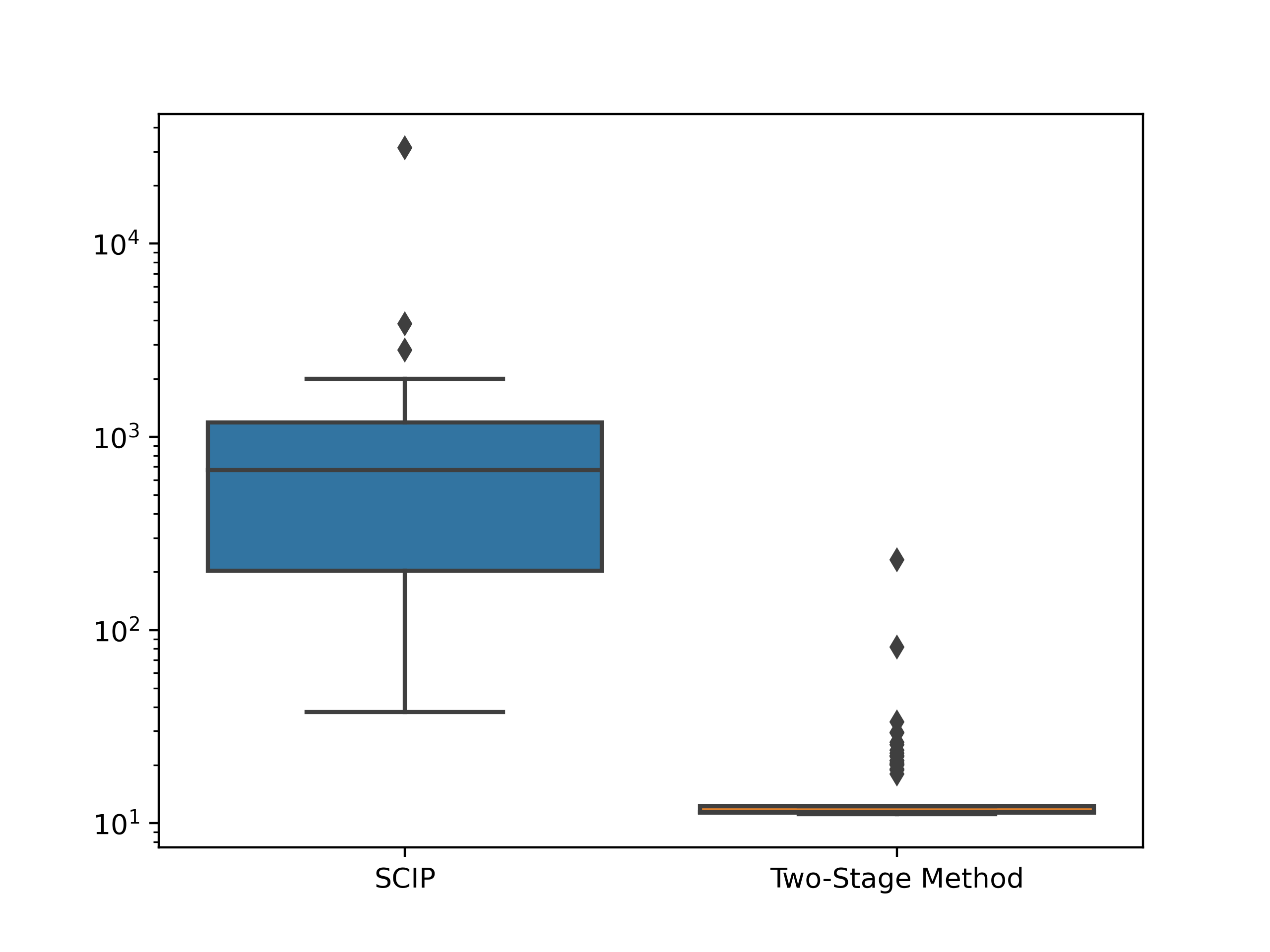}
    \caption{The comparison of solving time between the SCIP and our two-stage method among test data samples on Case 18.}
    \label{fig: time}
\end{figure}

We compute the average objective values, $C_{ev}$, $C_g$ among test data samples and summarize them in Table \ref{tab: optimality and efficiency}. we have an average gap of $0.019\%$ and a time reduction of $98.460\%$ on Case 18 and an average gap of $0.007\%$ and a time reduction of $98.998\%$ on Case 69. These illustrate that our two-stage method has an excellent performance in solving quality and speed, and can effectively address the computational challenges of the mixed-integer nonlinear coordination of OPF and V2G scheduling. Figure \ref{fig: time} demonstrates the time to solve these MINP problems with our two-stage method and directly with SCIP on Case 18. It can be seen that our method can generally reduce the solving time.

\subsubsection{Sensitivity Analysis} 
\begin{table}[h!]
\begin{center}
\caption{Sensitivity Analysis on Case 18}
\label{tab: Sensitivity Analysis case 18}
\begin{tabular}{cccc}
\hline
$\Delta$ & $0$ & $1$ & $2$ \\
\hline
Feasibility Retio (\%)  & 97.531    & 98.765  & 100\\
Time (s)                & 7.315     & 15.054  & 17.578\\
\hline 
\end{tabular}
\end{center}
\end{table}
The size of the trust region also determines the performance of the two-stage method. For instance, when $\Delta=0$, it is equivalent to rounding and fixing certain binary variables. However, the discrepancies between $\bar{\mathbf{y}}_\mathcal{S}$ and the optimal solution $\mathbf{y}^\star_\mathcal{S}$ will lead to suboptimal or even infeasible solutions. In Table \ref{tab: Sensitivity Analysis case 18}, we conduct the sensitivity analysis for different $\Delta$ values within $\{0,1,2\}$ on Case 18. One can notice that when $\Delta \in \{0, 1\}$, it cannot guarantee zero-violation on constraints among all data samples. Whereas, as $\Delta$ increases, the security of operations in the smart grid will be ensured. In addition, the solving time also has a positive correlation with the value of $\Delta$. 

\begin{table}[ht!]
\begin{center}
\caption{Sensitivity Analysis on Case 69}
\label{tab: Sensitivity Analysis case 69}
\begin{tabular}{ccc}
\hline 
$\Delta$ & 0 & 5\\
\hline 
Gap (\%) & 0.051 & 0.007   \\
Time (s) & 71.648 & 89.939 \\
\hline
\end{tabular}
\end{center}
\end{table}

In Table \ref{tab: Sensitivity Analysis case 69}, we conduct the sensitivity analysis on Case 69 with $\Delta$ of 0 and 5. Since both of them can ensure feasibility, we focus on the gap of the objective value. It can be seen that the larger the $\Delta$, the smaller the gap from the optimum.

\section{Conclusion}
This paper presented a novel two-stage optimization method integrating the difference of convex and trust region optimization techniques to address the computationally challenging MINP problem of coordinating optimal power flow with vehicle-to-grid scheduling. In stage one, this method utilized the DC algorithm to relax integrality constraints and reformulated MINP into a sequence of path-following continuous programs. Although the DC algorithm efficiently improved the solving speed, the solution may remain infeasible. Consequently, we proposed a trust region optimization to ensure feasible and high-quality solutions. The trust region was constructed according to the outcomes of the DC algorithm and we searched for the best feasible solution within the region. Simulation results substantiated that our method significantly enhances computational efficiency while achieving near-optimal solutions. In future work, we will extend this proposed method to distributed scenarios and address the uncertainty of EVs.

% In practice, V2G control coordinating with the power system needs to meet shorten response time, and the large-scale EVs connected to the grid also means exponential growth in computing challenges. By applying our two-stage approach, grid systems can more efficiently deliver solutions that have the potential to be applied to large-scale control scenarios with real-time requirements.

% Generated by IEEEtran.bst, version: 1.14 (2015/08/26)

\vspace{12pt}

\end{document}